\documentclass [12 pt]{article}
\def\be{\begin{equation}}
\def\eea{\end{eqnarray}}
\def\bea{\begin{eqnarray}}
\def\ee{\end{equation}}

\usepackage[psamsfonts]{amssymb}
\usepackage{amsmath}
\author{M. Amooshahi$^{1}$ \footnote{amooshahi@sci.ui.ac.ir}
\\ $^{1}$ {\small Faculty of science, University of Isfahan ,Hezar Jarib Ave.,
Isfahan,Iran}}
\title{Canonical quantization of electromagnetic field in the presence of absorbing bi-anisotropic multilayer
magnetodielectric media}
\begin{document}
\maketitle
\begin{abstract}
A bi-anisotropic magnetodielectric medium is modeled by two
independent set of three dimensional harmonic oscillators .A fully
canonical quantization of electromagnetic field is demonstrated in
the presence of a bi-anisotropic magnetodielectric medium. The
electric and magnetic polarization fields of the medium are obtained
in terms of the dynamical variable modeling the medium. The
Heisenberg equations of the system are solved for a multilayer
bi-anisotropic magnetodielectric medium.
\end{abstract}
\section{Introduction}
It is well known that the quantum properties of electromagnetic
field can be influenced by the presence of magnetodielectric
media.Typical examples are the casimir forces \cite{1}-\cite{4} and
the modification
of the spontaneous emission rate of excited atoms in the presence of polarizable and magnetizable media \cite{5}-\cite{16}.\\
There are mainly three approaches to quantize electromagnetic field
in the presence of magnetodielectric bodies. One approach is the
damped polarization model \cite{17}-\cite{19.1}. In this method the
electric polarization of the medium is represented by a quantum
field and absorptivity character of the dielectric medium is
described by interaction between the polarization field with a heat
bath containing a continuous set of harmonic oscillators. In this
approach a canonical quantization is  formulated for electromagnetic
field and the polarizable medium. The dielectric function of the
medium is obtained in terms of the coupling function of the heat
bath and the electric polarization field such that it satisfy the
Kramers-Kronig relations.\\
In a second approach, to quantize electromagnetic field in the
presence of magnetodielectric media, by adding the noise electric
and magnetic polarization densities to the classical constitutive
equations of the medium, these equations is taken as the definitions
of electric and magnetic polarization operators \cite{20}-\cite{27}.
The noise polarizations are related to two independent sets of
bosonic operators. Then, combination of the Maxwell equations and
the constitutive equations in frequency domain, gives the
electromagnetic field operators in terms of the noise polarizations
and the classical Green tensor. Suitable commutation relations are
imposed on the bosonic operators such that the commutation relations
between electromagnetic field operators become identical with those
in free space.\\
In a third scheme a fully canonical quantization of electromagnetic
field has been introduced in the presence of an anisotropic
polarizable and magnetizable medium\cite{28},\cite{29}. In this
method the medium is modeled by two independent reservoirs. Each
reservoir contains a continuum of three dimensional harmonic
oscillators.  The two reservoirs  describe polarizability a and
magnetizability of the medium. In contrast of the damped
polarization model, introduced above, in this approach the electric
and magnetic polarization fields of the medium do not need to appear
explicitly in the Lagrangian of the total system as a part of the
degrees of freedom of the medium. The contribution of the medium in
the Lagrangian of the total system is related only to the two
reservoirs and these reservoirs  completely constitute the degrees
of freedom of the medium.\\
In the present paper, we generalize the third  approach \cite{28}-\cite{29.1}for
bi-anisotropic magnetodielectric  media and then, the Maxwell
equations in the Heisenberg picture are solved                                                                                                                             exactly for a multilayer
medium.
\section{The Lagrangian of the system} In this section  we
generalize the canonical quantization method in the reference
\cite{28} for bi-anisotropic media. In this approach  the
magneto-dielectric is modeled by two continuum  collection  of three
dimensional harmonic oscillators and the total lagrangian  is
proposed  as
\begin{equation}\label{c1}
L(t)=\int d^3r [ \pounds_{s}+ \pounds_{em}+\pounds_{int}].
 \end{equation}
The first term  in the integrand is the contribution related to the
medium and is written as
\begin{equation}\label{c2}
\pounds_{s}= \int _0^\infty d\omega  \sum_{i=1}^2\left[ \frac{1}{2}\
 \dot{{\bf X}}_{\omega}^{( i)}({\bf r},t) \cdot \dot{{\bf X}}_{\omega}^{( i)}({\bf r},t)
 -\frac{1}{2}\ \omega^2\
 {\bf X}_{\omega}^{( i)}({\bf r},t)\cdot  {\bf X}_{\omega}^{( i)}({\bf r},t)\right]
\end{equation}
where  $ {\bf X}^{(i)}_\omega({\bf r},t)$ for $ i=1,2$ is the
dynamical variables of the harmonic oscillator labeled by the
continuous parameter $\omega$. The second part in in the integrand
(\ref{c1}) is the Lagrangian density of the electromagnetic field
which is written as usual way as
\begin{equation}\label{c3}
\pounds_{em}=   \frac{1}{2}\varepsilon _0 {\bf E}^2-\frac{{\bf
B}^2}{2\mu_0},
\end{equation}
where ${\bf E}=- \frac{\partial{\bf A}}{\partial
t}-\vec{\nabla\varphi}$ , ${\bf B}=\nabla\times{\bf A}$  are
electric and magnetic fields respectively and  ${\bf A}$ and
$\varphi $ are the vector and the scalar potentials. The  last term
$\pounds_{int}$ in (\ref{c1}) is describing the interaction between
 the bi-anisotropic magnetodielectric medium and the electromagnetic field and is
as follows
\begin{eqnarray}\label{c4}
\pounds_{int}&=& \int _0^\infty d \omega \sum_{i=1}^2\
\left(\sum_{m,n=1}^3 f^{mn}_{i}( \omega , {\bf r})E_m({\bf r},t)
X_{\omega n }^{(i)}({\bf r},t)\right)
 \nonumber\\
&+&\int _0^\infty d \omega \sum_{i=1}^2\ \left(\sum_{m,n=1}^3
g^{mn}_{i}( \omega , {\bf r}) B_m({\bf r},t)X_{\omega n }^{(i)}({\bf
r},t)\right)
\end{eqnarray}
where ${\bf f}_{i}, {\bf g}_{i}, i=1,2 $ are the coupling tensors of
the second rank between the electromagnetic field and the medium.
The interaction part (\ref{c4}) is the generalization of the
Lagrangian that has previously been used for anisotropic spatially
dispersion magneto-dielectric media\cite{28}.\\
As mentioned above, in contrast of the damped polarization method,
in this scheme the electric and magnetic polarization fields are not
appeared explicitly in the Lagrangian density of the total system
and the contribution of the medium in the lagrangian merely is
$\pounds_s$ containing two independent reservoirs.
\section{Classical Euler-Lagrange equations}
 Using the Lagrangian
density (\ref{c2})-(\ref{c4}) it can be shown that the classical
Euler-Lagrange equations for the scalar and vector potentials
$\varphi$ and $ {\bf A}$ leads to the Gauss' law and the Maxwell
equation, respectively as
\begin{eqnarray}\label{c5}
\nabla\cdot(\varepsilon_0{\bf E}+{\bf P})=0
\end{eqnarray}
\begin{eqnarray}\label{c5.1}
\nabla\times(\frac{{\bf B}}{\mu_0}-{\bf
M})&=&\frac{\partial}{\partial t}(\varepsilon_0{\bf E}+{\bf P}),
\end{eqnarray}
where $ {\bf P}$ and ${\bf M}$ are the electric and magnetic
polarization densities of the medium which are defined in the terms
of the oscillators modeling the medium as
\begin{eqnarray}\label{c6}
{\bf P}({\bf r},t)&=&\int_0^\infty d\omega \sum_{i=1}^2\ {\bf
f}_{i}(\omega,{\bf r})\cdot\ {\bf X}_{\omega}^{( i)}({\bf
r},t)\nonumber\\
{\bf M}({\bf r},t)&=&\int_0^\infty d\omega \sum_{i=1}^2\ {\bf
g}_{i}(\omega,{\bf r})\cdot\ {\bf X}_{\omega }^{(i)}({\bf r},t).
\end{eqnarray}
Also the Euler-Lagrange equations for the dynamical variables ${\bf
X}^{(i)}_\omega,\ i=1,2$ leads to
\begin{eqnarray}\label{c7}
\ddot{{\bf X}}_{\omega}^{( i)}({\bf r},t)+\omega^2\ {\bf
X}_{\omega}^{( i)}({\bf r},t)= {\bf f}_{i}^t(\omega,{\bf
r})\cdot{\bf E}({\bf r},t)+ {\bf g}_{i}^t(\omega,{\bf r})\cdot{\bf
B}({\bf r},t).
\end{eqnarray}
where the superscript  $t$  imply  the transposition. The formally
solution of the differential equations ( \ref{c7}) can be written as
\begin{eqnarray}\label{c7.1}
{\bf X}^{(i)}_\omega({\bf r},t)&=&{\bf X}^{(i)}_\omega({\bf
r},0)\cos\omega t+\frac{\sin\omega t}{\omega}\dot{{\bf
X}}^{(i)}_\omega({\bf r},0)\nonumber\\
&+&\int_0^t dt'\frac{\sin \omega(t-t')}{\omega}\left[{\bf
f}^t_i(\omega,{\bf r})\cdot{\bf E}({\bf r},t')+{\bf
g}^t_i(\omega,{\bf r})\cdot{\bf B}({\bf r},t')\right]
\end{eqnarray}
If one  substitute the solutions (\ref{c7.1}) into the definitions
(\ref{c6}), the constitutive relations of the bi-anisotropic
magneto-dielectric medium are obtained as
\begin{eqnarray}\label{c8}
{\bf P}({\bf r},t)&=&{\bf P}_N({\bf r},t)+\int_0^{|t|} dt'\left[
{\bf\chi}^{(1)}({\bf r},|t|-t')\cdot {\bf E}({\bf r}, \pm
t')+{\bf\chi}^{(2)}({\bf r},|t|-t')\cdot {\bf B}({\bf
r},\pm t')\right]\nonumber\\
{\bf M}({\bf r},t)&=&{\bf M}_N({\bf r},t)+\int_0^{|t|} dt'\left[
{\bf\chi}^{(3)}({\bf r},|t|-t')\cdot {\bf E}({\bf r}, \pm
t')+{\bf\chi}^{(4)}({\bf r},|t|-t')\cdot {\bf B}({\bf
r},\pm t')\right]\nonumber\\
&&
\end{eqnarray}
where the upper (lower ) sign  is for $t>0$ ( $t<0$) and the tensors
$ \chi^{(i)},\ i=1,2,3,4 $ are the susceptibility tensors of the
medium and are defined in terms of the coupling tensors of the
electromagnetic field and the medium as the following
\begin{eqnarray}\label{c9}
&&\chi^{(1)}({\bf r},t)=\int_0^\infty d\omega\ \frac{\sin \omega
t}{\omega}\left[ {\bf f}_{1}(\omega,{\bf r})\cdot {\bf
f}_{1}^t(\omega,{\bf r})+{\bf f}_{2}(\omega,{\bf r})\cdot{\bf
f}_{2}^t(\omega,{\bf r})\right]\nonumber\\
&&\chi^{(4)}({\bf r},t)=\int_0^\infty d\omega\ \frac{\sin \omega
t}{\omega}\left[ {\bf g}_{1}(\omega,{\bf r})\cdot {\bf
g}_{1}^t(\omega,{\bf r})+{\bf g}_{2}(\omega,{\bf r})\cdot{\bf
g}_{2}^t(\omega,{\bf r})\right]\nonumber\\
&&\chi^{(2)}({\bf r},t)=(\chi^{(3)})^t({\bf
r},t)=\nonumber\\
&&\int_0^\infty d\omega\ \frac{\sin \omega t}{\omega}\left[ {\bf
f}_{1}(\omega,{\bf r})\cdot {\bf g}_{1}^t(\omega,{\bf r})+{\bf
f}_{2}(\omega,{\bf r})\cdot{\bf
g}_{2}^t(\omega,{\bf r})\right]\nonumber\\
&&
\end{eqnarray}
In the constitutive relations (\ref{c8}) the fields $ {\bf P}_N$ and
$ {\bf M}_N$ are the noise polarization densities  which can be
written as
\begin{eqnarray}\label{c10}
{\bf P}_N({\bf r},t)&=&\int_0^\infty d\omega \ \sum_{i=1}^2\left[
{\bf f}_{i}(\omega,{\bf r})\cdot\left({\bf X}_{\omega}^{( i)}({\bf
r},0)\cos\omega t+\dot{{\bf X}}_{\omega}^{( i)}({\bf r},0)\frac{\sin
\omega t}{\omega}\right)\right]\nonumber\\
{\bf M}_N({\bf r},t)&=&\int_0^\infty d\omega \ \sum_{i=1}^2\left[
{\bf g}_{i}(\omega,{\bf r})\cdot\left({\bf X}_{\omega }^{(i)}({\bf
r},0)\cos\omega t+\dot{{\bf X}}_{\omega}^{( i)}({\bf r},0)\frac{\sin
\omega t}{\omega}\right)\right]\nonumber\\
&&
\end{eqnarray}
An anisotropic magnetodielectric medium is  the medium that its
constitutive relation are as
\begin{eqnarray}\label{c10.1}
{\bf P}({\bf r},t)&=&{\bf P}_N({\bf r},t)+\int_0^{|t|} dt'
{\bf\chi}_e({\bf r},|t|-t')\cdot {\bf E}({\bf r}, \pm
t')\nonumber\\
{\bf M}({\bf r},t)&=&{\bf M}_N({\bf r},t)+\int_0^{|t|} dt'
{\bf\chi}_m({\bf r},|t|-t') {\bf B}({\bf
r},\pm t')\nonumber\\
&&
\end{eqnarray}
where $ \chi_e $ and $\chi_m$ are the electric and magnetic
susceptibilities tensors. In order to obtain Eq.(\ref{c10.1}), in
the Heisenberg picture, in the Lagrangian density one set of the
oscillator modeling the medium should be coupled to the electric
field and the other should be coupled to the magnetic field,
separately  \cite{28}. A bi-anisotropic magnetodielectric medium is
meant the medium that the constitutive relations are as
(\ref{c8}),where $ \chi^{(i)}\ i=1,2,3,4$ are tensors of the second
rank. that is, in bi-anisotropic magnetodielectric medium the
electric and magnetic polarization densities are dependent on both
the electric and magnetic field. In order to obtain Eqs. (\ref{c8}),
in the Heisenberg picture,  both collections of oscillators modeling
the medium should be coupled with both electric and magnetic field.
The oscillators modeling the medium describe the absorption of the
energy of the electromagnetic field by the medium due to temporal
dispersive property. Also as it is clear from Eqs.(\ref{c6}) the
electric and magnetic polarization densities are defined in terms of
the oscillators modeling the medium. Therefore the two collections
of the oscillators describe polarizability, magnetizability and the
absorption due to temporal dispersive property of the
magnetodielectric medium. In fact this scheme of quantization is a
generalization of the Caldeira-Legget model for dissipative system
\cite{30} and \cite{31}. In the Caldeira-Legget model the absorbing
environment of a main dissipative system is modeled by a set of
harmonic oscillators. The oscillators describe the absorption of the
energy of the main system by the environment. In the present
quantization  our main dissipative system is the electromagnetic
field and the magnetodielectric medium play the role of the
absorbing environment. Accordingly the oscillators modeling the
medium describe the absorption of the energy by the absorbing
magnetodielectric medium.
\section{Canonical Quantization}
To have a canonical quantization of the combined system, that is the
electromagnetic field and the medium, we are confronted with the
usual problem that the scalar potential $\varphi$ does not possess a
canonically conjugate variable, since the lagrangian density of the
total system does not contain the time derivative of the the scalar
potential. To overcome  this , as the usual way,  the Gauss' law
(\ref{c5}) together with the Coulomb gauge $\nabla\cdot{\bf A}=0$
can be used to eliminate the extra degree's of freedom from the
Lagrangian of the total system. Applying the Coulomb gauge
$\nabla\cdot{\bf A}=0$ in the Gauss' law (\ref{c5})  the scalar
potential can be chosen as
\begin{eqnarray}\label{c11}
\varphi({\bf r},t)=\frac{1}{4\pi\varepsilon_0}\int d^3r'\
\frac{\nabla'\cdot{\bf P}({\bf r'},t)}{|{\bf r}-{\bf r'}|}
\end{eqnarray}
 substitution of the constraint (\ref{c11})  into the Lagrangian (\ref{c1})-(\ref{c4}) and apply
several integration by parts, for the total Lagrangian of the
system, we have
\begin{eqnarray}\label{c12}
L(t)&=&\int d^3r \left[ \frac{1}{2}\varepsilon_0(\frac{\partial{\bf
A}}{\partial t})^2-\frac{(\nabla\times{\bf
A})^2}{2\mu_0}-\frac{\partial{\bf A}}{\partial t}\cdot{\bf
P}^T+(\nabla\times{\bf A})\cdot M\right]\nonumber\\
&+&\int_0^\infty d\omega\ \sum_{i=1}^2\left[\frac{1}{2}\
 (\dot{{\bf X}}^{(i)}_\omega)^2-\frac{1}{2}\ \omega^2\
 ({\bf X}^{(i)}_\omega)^2 \right]\nonumber\\
 &-&\frac{1}{4\pi\varepsilon_0}\int d^3r\ \int d^3r'\
 \frac{\nabla\cdot{\bf P}({\bf r},t)\ \nabla'\cdot{\bf P}({\bf r'},t)}{|{\bf r}-{\bf r'}|}
\end{eqnarray}
where the polarization fields ${\bf P} , {\bf M}$ are given by
(\ref{c6}) and $ {\bf P}^T$ is the transverse component of the
electric polarization field ${\bf P}$ defined by
\begin{eqnarray}\label{c13}
P^T_i({\bf r},t)=\sum_{j=1}^3\int d^3r'\ \delta^T_{ij}({\bf r},{\bf
r'})\ P_j({\bf r'},t)
\end{eqnarray}
where
\begin{eqnarray}\label{c14}
\delta^T_{ij}({\bf r},{\bf r'})=\delta_{ij}\delta({\bf r}-{\bf
r'})-\frac{1}{4\pi}\frac{\partial}{\partial x_i}\frac{1}{|{\bf
r}-{\bf r'}|}\ \frac{\partial}{\partial x'_j}
\end{eqnarray}
is the transverse Dirac operator. Now  the Lagrangian (\ref{c12})
possess only the vector potential  as the dynamical variable of the
electromagnetic field. One can compute the conjugate variables of
the fields $ {\bf A}, {\bf X}^{(1)}$ and ${\bf X}^{(2)}$
respectively  as
\begin{eqnarray}\label{c15}
-{\bf D}&=&\frac{\delta L}{\delta \dot{{\bf A}}}=\varepsilon_0
\dot{{\bf A}}-{\bf P}^T\\
{\bf Q}^{(i)}_\omega&=&\frac{\delta L}{\delta \dot{{\bf
X}}^{(i)}_\omega}=\dot{{\bf X}}^{(i)}_\omega  \hspace {2.00 cm}
i=1,2
\end{eqnarray}
Having the conjugate variables of the combined system, as the
standard fashion, we postulate the following  canonical commutation
relations  on the cartesian components of the field operators of the
system
\begin{eqnarray}\label{c16}
[A_i({\bf r},t)\ , \ -D_j({\bf r'},t)]=i\hbar \delta^T_{ij}({\bf
r},{\bf r'})
\end{eqnarray}
\begin{eqnarray}\label{c16.1}
 &&[{\bf X}^{(i)}_\omega({\bf r},t)\ ,\ {\bf
Q}^{(j)}_{\omega'}({\bf r'},t)]= i \hbar I
\delta_{ij}\delta(\omega-\omega')\delta({\bf r}-{\bf r'})
\end{eqnarray}
where $I$ is the unit matrix. Using These commutation relations  and
the Hamiltonian
\begin{eqnarray}\label{c17}
H(t)&=&\int d^3r \left[ \int_0^\infty d\omega \sum_{i=1}^2 {\bf
Q}^{(i)}_\omega \cdot \dot{{\bf X}}^{(i)}_\omega-{\bf D}\cdot
\dot{{\bf
A}}\right]-L(t)\nonumber\\
&=&\int d^3 r\left[ \frac{({\bf D}-{\bf
P}^T)^2}{2\varepsilon_0}+\frac{(\nabla\times{\bf
A})^2}{2\mu_0}-(\nabla\times{\bf A})\cdot
M\right]\nonumber\\
&+&\int d^3r \left[\sum_{i=1}^2[\frac{1}{2}({\bf
Q}^{(i)}_\omega)^2+\frac{1}{2}\omega^2({\bf
X}^{(i)})^2]\right]\nonumber\\
&+&\frac{1}{4\pi\varepsilon_0}\int d^3r\ \int d^3r'\
 \frac{\nabla\cdot{\bf P}({\bf r},t)\ \nabla'\cdot{\bf P}({\bf r'},t)}{|{\bf r}-{\bf r'}|}
\end{eqnarray}
If one write the Heisenberg equation for the vector potential ${\bf
A}$ the Maxwell equation (\ref{c5.1}) can be reobtained in the
Heisenberg picture. Since it has been assumed that the volume
contained the electromagnetic field and the medium is the unbounded
space, we can expand the field operators of the system in terms of
the plane waves. In accordance to the Coulomb gauge, the expansions
of the vector potential and the displacement fields are
\begin{eqnarray}\label{c18}
{\bf A}({\bf r},t)&=&\sum_{\lambda=1}^2\ \int d^3k\
\sqrt{\frac{\hbar}{2(2\pi)^3\omega_{{\bf k}}}}[a_{{\bf
k}\lambda}(t)\ e^{i {\bf k}\cdot{\bf r}}+a^\dag_{{\bf k}\lambda}(t)
e^{-i {\bf k}\cdot{\bf r}}]{\bf e}_{{\bf k}\lambda} \\
{\bf D}({\bf r},t)&=&-i\sum_{\lambda=1}^2\ \int d^3k\
\sqrt{\frac{\hbar \omega_{{\bf k}}}{2(2\pi)^3}}[a_{{\bf
k}\lambda}^\dag(t)\ e^{-i {\bf k}\cdot{\bf r}}-a_{{\bf k}\lambda}(t)
e^{i {\bf k}\cdot{\bf r}}]{\bf e}_{{\bf k}\lambda}
\end{eqnarray}
where $\omega_{{\bf k}}=c|{\bf k}|$ and ${\bf e}_{{\bf k}\lambda}\
\lambda=1,2$ together with ${\bf e}_{{\bf k} 3}=\frac{{\bf k}}{|{\bf
k}|}$ constitute an orthonormal basis. Also the expansions of the
field operators of the magneto-dielectric medium are as follows
\begin{eqnarray}\label{c19}
{\bf X}^{(i)}_\omega({\bf r},t)&=&\sum_{\nu=1}^3\ \int d^3k\
\sqrt{\frac{\hbar}{2(2\pi)^3\omega_{{\bf k}}}}[b_{{\bf
k}\nu}^{(i)}(\omega,t)\ e^{i {\bf k}\cdot{\bf
r}}+(b^{(i)})^\dag_{{\bf k}\nu}(\omega,t) e^{-i {\bf k}\cdot{\bf
r}}]{\bf e}_{{\bf k}\nu}\nonumber\\
&&
\end{eqnarray}
\begin{eqnarray}\label{c20}
{\bf Q}^{(i)}_\omega({\bf r},t)&=&-i\sum_{\nu=1}^3\ \int d^3k\
\sqrt{\frac{\hbar \omega_{{\bf k}}}{2(2\pi)^3}}[b_{{\bf
k}\nu}^{(i)}(\omega,t)\ e^{i {\bf k}\cdot{\bf
r}}-(b^{(i)})^\dag_{{\bf k}\nu}(\omega,t) e^{-i {\bf k}\cdot{\bf
r}}]{\bf e}_{{\bf k}\nu}\nonumber\\
&&
\end{eqnarray}
Regarding to the commutation relations (\ref{c16}) and
(\ref{c16.1}), it is clear that  the ladder operators of the system
satisfy the commutation rules
\begin{eqnarray}\label{c21}
[a_{{\bf k}\lambda}(t)\ ,\ a_{{\bf
k'}\lambda'}^\dag(t)]=\delta_{\lambda\lambda'}\delta({\bf k}-{\bf
k'})
\end{eqnarray}
\begin{eqnarray}\label{c22}
[b_{{\bf k}\nu}^{(i)}(\omega,t)\ ,\ (b_{{\bf
k'}\nu'}^{(j)})^\dag(\omega',t)]=\delta_{ij}\delta_{\nu\nu'}\delta(\omega-\omega')\delta({\bf
k}-{\bf k'})
\end{eqnarray}
\section{multilayer magneto-dielectric media}
The coupled constitutive relations (\ref{c8}) and the Maxwell
equations can be solved using the forward and backward Laplace
transformation\cite{19}. Let us denote the backward and forward
Laplace transformation of a time dependent operator $O(t)$,
respectively by $O^b(s)$ and $O^f(s)$. The operator valued functions
$O^b(s)$ and $O^f(s)$ are defined as
\begin{eqnarray}\label{c23}
O^b(s)=\int_0^\infty O(t)e^{-s t} dt\hspace{2.00
cm}O^f(s)=\int_0^\infty O(-t)e^{-s t} dt
\end{eqnarray}
For a bi-anisotropic medium  it is useful to express the Laplace
transformations of the polarization fields ${\bf P}(t)$ and ${\bf
M}(t)$ in terms of the Laplace transformations of the electric and
magnetic fields ${\bf E}(t)$ and ${\bf H}(t)$. Using the
constitutive relations (\ref{c8}) we can write
\begin{eqnarray}\label{c24}
{\bf P}^{f,b}({\bf r},s)=({\bf P'}_N)^{f,b}({\bf
r},s)+\eta^{(1)}({\bf r},s){\bf E}^{f,b}({\bf r},s)+\eta^{(2)}({\bf
r} ,s) {\bf H}^{f,b}({\bf r},s)
\end{eqnarray}
\begin{eqnarray}\label{c25}
{\bf M}^{f,b}({\bf r},s)=({\bf M'}_N)^{f,b}({\bf
r},s)+\eta^{(3)}({\bf r},s){\bf E}^{f,b}({\bf r},s)+\eta^{(4)}({\bf
r} ,s) {\bf H}^{f,b}({\bf r},s)
\end{eqnarray}
where the superscripts $f$ and $b$ in the left hand are applied
correspondingly for superscripts $f$ and $b$ in the right hand of
this equation , respectively and
\begin{eqnarray}\label{c26}
\eta^{(1)}({\bf r},s)=\chi^{(1)}({\bf r},s)+\mu_0\chi^{(2)}({\bf
r},s)[I-\mu_0\chi^{(3)}({\bf r},s)]^{-1}\chi^{(4)}({\bf r},s)
\end{eqnarray}
\begin{eqnarray}\label{c27}
\eta^{(2)}({\bf r},s)=\mu_0\chi^{(2)}({\bf
r},s)[I-\mu_0\chi^{(3)}({\bf r},s)]^{-1}
\end{eqnarray}
\begin{eqnarray}\label{c28}
\eta^{(3)}({\bf r},s)=[I-\mu_0\chi^{(3)}({\bf
r},s)]^{-1}\chi^{(4)}({\bf r},s)
\end{eqnarray}
\begin{eqnarray}\label{c29}
\eta^{(4)}({\bf r},s)=\mu_0\chi^{(3)}({\bf
r},s)[I-\mu_0\chi^{(3)}]^{-1}({\bf r},s)
\end{eqnarray}
\begin{eqnarray}\label{c30}
({\bf P'}_N)^{f,b}={\bf P}^{f,b}_N({\bf r},s)+\mu_0\chi^{(2)}({\bf
r},s)[I-\mu_0\chi^{(3)}({\bf r},s)]^{-1}{\bf M}^{f,b}_N({\bf r},s)
\end{eqnarray}
\begin{eqnarray}\label{c31}
({\bf M'}_N)^{f,b}({\bf r},s)=[I-\mu_0\chi^{(3)}({\bf
r},s)]^{-1}{\bf M}^{f,b}_N({\bf r},s)
\end{eqnarray}
where $I$ is the unit tensor. It should be noted that since the
susceptibility tensors are identically zero for $t\leq0$, their
backward Laplace transformation vanishes. Therefore only the forward
Laplace transformation of the tensors are appeared in
(\ref{c24})-(\ref{c31}) that we have denoted them without
superscript. Now combination of the constitutive relations
(\ref{c24}) , (\ref{c25}) with the Laplace transformations of
Faraday's law and the Maxwell equation (\ref{c5.1}) gives the
following six coupled differential equations for the cartesian
components of the fields {\bf E} and {\bf H}
\begin{eqnarray}\label{c32}
&&\left[\begin{array}{cc}
        T^{f,b}({\bf r},s) & Y^{f,b}({\bf r},s) \\
        Z^{f,b}({\bf r},s) & W^{f.b}({\bf r},s) \\
      \end{array}\right]\left[\begin{array}{c}
                                {\bf E}^{f,b}({\bf r},s) \\
                                {\bf H}^{f,b}({\bf r},s) \\
                              \end{array}\right]=J^{f,b}({\bf r},s)
\end{eqnarray}
where
\begin{eqnarray}\label{c32.1}
                        J^{f,b}({\bf r},s)=\left[\begin{array}{c}
                                      \mp\mu_0s({\bf M'})_N^{f,b}({\bf r},s)\pm{\bf B}({\bf r},0) \\
                                     \pm s({\bf P'})_N^{f,b}({\bf r},s)\mp{\bf D}({\bf r},0)  \\
                                    \end{array}\right]
\end{eqnarray}
and
\begin{eqnarray}\label{c33}
&&T^{f,b}_{ij}=\sum_{k=1}^3 \epsilon_{ikj}\partial_k\pm\mu_0 s
\eta^{(3)}_{ij}\hspace{1.50 cm}Y^{f,b}=\pm\mu_0
s(I+\eta^{(4)})\nonumber\\
&&W^{f,b}_{ij}=\sum_{k=1}^3 \epsilon_{ikj}\partial_k \mp  s
\eta^{(2)}_{ij}\hspace{1.50 cm}Z^{f,b}= \mp
s(\varepsilon_0I+\eta^{(1)})
\end{eqnarray}
 where  $ \epsilon_{ikj}$ is three
dimensional Levi-cibita symbol and the upper (lower) sign  is
applied for the forward (backward) Laplace transformation.
Hereafter, the upper (lower) sign in any equation is used for the
forward (backward) Laplace transformation and the superscripts $f$
and $b$ in the left hand of any equation are applied
correspondingly for the superscripts $f$ and $b$ in the right hand
of that equation, respectively. For a multilayer medium, that is
when the susceptibility tensors are independent of the coordinates
$x, y$ and are piecewisely constant with respect to the coordinate
$z$, The symmetry of configuration enables us to convert the coupled
partial differential equations (\ref{c32}) into a set of first order
ordinary differential equations with respect to the coordinate $z$.
This can be achieved by expressing each field operators of the total
system in terms of two dimensional Fourier transform with respect to
coordinates $x$ and $y$. For example for the electric field this
becomes
\begin{eqnarray}\label{c34}
{\bf E}^{(f,b)}({\bf r},s)=\frac{1}{2\pi}\int_{-\infty}^{+\infty}
d^2k\  \underline{\bf{E}}^{f,b}({\bf k}^\|,s,z)e^{ \pm i{\bf
k}^\|\cdot {\bf r}^\|}
\end{eqnarray}
where ${\bf k}^\|=k_x{\bf i}+k_y{\bf j}$ and ${\bf r}^\|=x{\bf
i}+y{\bf j}$. Applying this transformation into Eqs.(\ref{c32}),
These equations are reduced to
\begin{eqnarray}\label{c35}
\left[\begin{array}{cc}
        \underline{T}^{f,b} & \underline{Y}^{f,b} \\
        \underline{Z}^{f,b} & \underline{W}^{f.b} \\
      \end{array}\right]\left[\begin{array}{c}
                                {\bf \underline{E}}^{f,b} \\
                                {\bf \underline{H}}^{f,b} \\
                              \end{array}\right]=\underline{J}^{f,b}({\bf
                            k}^\|,z,s)
\end{eqnarray}
where now
\begin{eqnarray}\label{c36}
&&\underline{T}^{f,b}_{12}=-\frac{\partial}{\partial z}\pm \mu_0 s
\eta^{(3)}_{12}(z,s)\hspace{1.00
cm}\underline{T}^{f,b}_{21}=\frac{\partial}{\partial z}\pm \mu_0 s
\eta^{(3)}_{21}(z,s)\nonumber\\
&&\underline{T}^{f,b}_{13}=\pm ik_y\pm \mu_0 s
\eta^{(3)}_{13}(z,s)\hspace{1.00 cm}\underline{T}^{f,b}_{31}=\mp
ik_y\pm \mu_0 s
\eta^{(3)}_{31}(z,s)\nonumber\\
&&\underline{T}^{f,b}_{23}=\mp ik_x\pm \mu_0 s
\eta^{(3)}_{23}(z,s)\hspace{1.00 cm}\underline{T}^{f,b}_{32}=\pm
ik_x\pm \mu_0 s
\eta^{(3)}_{32}(z,s)\nonumber\\
&&\underline{T}^{f,b}_{ii}=\pm\mu_0 s
\eta^{(3)}_{ii}(z,s)\hspace{2.00 cm} i=1,2,3
\end{eqnarray}
and
\begin{eqnarray}\label{c37}
&&\underline{W}^{f,b}_{12}=-\frac{\partial}{\partial z}\mp s
\eta^{(2)}_{12}(z,s)\hspace{1.00
cm}\underline{W}^{f,b}_{21}=\frac{\partial}{\partial z}\mp s
\eta^{(2)}_{21}(z,s)\nonumber\\
&&\underline{W}^{f,b}_{13}=\pm ik_y\mp s
\eta^{(2)}_{13}(z,s)\hspace{1.00 cm}\underline{W}^{f,b}_{31}= \mp
ik_y\mp s
\eta^{(2)}_{31}(z,s)\nonumber\\
&&\underline{W}^{f,b}_{23}= \mp ik_x\mp s
\eta^{(2)}_{23}(z,s)\hspace{1.00 cm}\underline{W}^{f,b}_{32}=\pm
ik_x\mp
s\eta^{(2)}_{32}(z,s)\nonumber\\
&&\underline{W}^{f,b}_{ii}=\mp s \eta^{(2)}_{ii}(z,s)\hspace{2.00
cm} i=1,2,3
\end{eqnarray}
and the definitions of  $\underline{Y}^{f,b}, \underline{Z}^{f,b}$
are the same as in (\ref{c33}). The form of the differential
equations frame (\ref{c35}) and the definitions
(\ref{c36}),(\ref{c37}) shows  that the third and the sixth equation
in (\ref{c35}) are algebraic equations. This can be used to
eliminate the $z$ components of the fields ${\bf
\underline{E}}^{f,b}$ and $\underline{H}^{f,b}$ in the first , the
second,the fourth and the fifth equation of the frame (\ref{c35}).
If we solve the third equation together with the sixth equation of
(\ref{c35}) for $\underline{E}^{f,b}_z$ and $\underline{H}^{f,b}_z$,
we obtain
\begin{eqnarray}\label{c38}
&&\underline{E}^{f,b}_z=\alpha_1\underline{E}^{f,b}_x+\beta_1\underline{E}^{f,b}_y+
\gamma_1\underline{H}^{f,b}_x+\delta_1\underline{H}^{f,b}_y \pm
\frac{\underline{W}_{33}\underline{J}^{f,b}
_3-\underline{Y}_{33}\underline{J}^{f,b}_6}{\underline{T}_{33}\underline{W}_{33}
-\underline{Z}_{33}\underline{Y}_{33}}\nonumber\\
&&\underline{H}^{f,b}_z=\alpha_2\underline{E}^{f,b}_x+\beta_2\underline{E}^{f,b}_y+
\gamma_2\underline{H}^{f,b}_x+\delta_2\underline{H}^{f,b}_y
\pm\frac{\underline{T}_{33}\underline{J}^{f,b}
_6-\underline{Z}_{33}\underline{J}^{f,b}_3}{\underline{T}_{33}\underline{W}_{33}
-\underline{Z}_{33}\underline{Y}_{33}}\nonumber\\
&&
\end{eqnarray}
where
\begin{eqnarray}\label{c39}
&&\alpha_1=\frac{\underline{Y}_{33}\underline{Z}_{31}
-\underline{W}_{33}\underline{T}_{31}}{\underline{T}_{33}\underline{W}_{33}
-\underline{Z}_{33}\underline{Y}_{33}},\
\beta_1=\frac{\underline{Y}_{33}\underline{Z}_{32}
-\underline{W}_{33}\underline{T}_{32}}{\underline{T}_{33}\underline{W}_{33}
-\underline{Z}_{33}\underline{Y}_{33}},\nonumber\\
\nonumber\\
&&\gamma_1=\frac{\underline{Y}_{33}\underline{W}_{31}
-\underline{W}_{33}\underline{Y}_{31}}{\underline{T}_{33}\underline{W}_{33}
-\underline{Z}_{33}\underline{Y}_{33}},\
\delta_1=\frac{\underline{Y}_{33}\underline{W}_{32}
-\underline{W}_{33}\underline{Y}_{32}}{\underline{T}_{33}\underline{W}_{33}
-\underline{Z}_{33}\underline{Y}_{33}}\nonumber\\
\nonumber\\
&&\alpha_2=\frac{\underline{Z}_{33}\underline{T}_{31}
-\underline{T}_{33}\underline{Z}_{31}}{\underline{T}_{33}\underline{W}_{33}
-\underline{Z}_{33}\underline{Y}_{33}},\
\beta_2=\frac{\underline{Z}_{33}\underline{T}_{32}
-\underline{T}_{33}\underline{Z}_{32}}{\underline{T}_{33}\underline{W}_{33}
-\underline{Z}_{33}\underline{Y}_{33}}\nonumber\\
\nonumber\\
&&\gamma_2=\frac{\underline{Z}_{33}\underline{Y}_{31}
-\underline{T}_{33}\underline{W}_{31}}{\underline{T}_{33}\underline{W}_{33}
-\underline{Z}_{33}\underline{Y}_{33}},\
\delta_2=\frac{\underline{Z}_{33}\underline{Y}_{32}
-\underline{T}_{33}\underline{W}_{32}}{\underline{T}_{33}\underline{W}_{33}
-\underline{Z}_{33}\underline{Y}_{33}}
\end{eqnarray}
and we have used the new notation
\begin{eqnarray}\label{c39.1}
&&\underline{T}_{13}=\underline{T}_{13}^f,\
\underline{T}_{31}=\underline{T}_{31}^f,\
\underline{T}_{23}=\underline{T}_{23}^f,\
\underline{T}_{32}=\underline{T}_{32}^f\nonumber\\
&&\underline{W}_{13}=\underline{W}_{13}^f,\
\underline{W}_{31}=\underline{W}_{31}^f,\
\underline{W}_{23}=\underline{W}_{23}^f,\
\underline{W}_{32}=\underline{W}_{32}^f\nonumber\\
&&\underline{Z}_{ij}=\underline{Z}_{ij}^f,\
\underline{Y}_{ij}=\underline{Y}_{ij}^f\hspace{1.50 cm}
i,j=1,2,3\nonumber\\
&&\underline{T}_{ii}=\underline{T}_{ii}^f\
,\underline{W}_{ii}=\underline{W}_{ii}\hspace{1.50 cm} i=1,2,3
\end{eqnarray}
Now substituting $\underline{E}^{f,b}_z$ and $\underline{H}^{f,b}_z$
from (\ref{c38}) into the first, the second,the fourth and the fifth
equation of the frame (\ref{c35}) give us a set of the coupled first
order differential equations for the other components of the fields
${\bf \underline{E}}^{f,b}$ and $\underline{H}^{f,b}$ as
\begin{eqnarray}\label{c40}
\frac{\partial}{\partial z}\Lambda^{f,b}({\bf
k}^\|,z,s)\pm\Theta({\bf k}^\|,z,s)\Lambda^{f,b}({\bf
k}^\|,z,s)=G^{f,b}({\bf k}^\|,z,s)
\end{eqnarray}
where
\begin{eqnarray}\label{c41}
\Lambda^{f,b}({\bf k}^\|,z,s)=\left[\begin{array}{c}
                                                  \underline{E}^{f,b}_x({\bf k}^\|,z,s) \\
                                                  \underline{E}^{f,b}_y({\bf k}^\|,z,s) \\
                                                  \underline{H}^{f,b}_x({\bf k}^\|,z,s) \\
                                                  \underline{H}^{f,b}_y({\bf k}^\|,z,s) \\
                                                \end{array}\right]
\end{eqnarray}
and the $\Theta ,\ G^{f,b}$ are $4\times4$ matrix and $4\times1$
matrix ,respectively, with scalar elements given by
\begin{eqnarray}\label{c41.1}
&&\Theta_{11}=\mu_0s\eta^{(3)}_{21}+\underline{T}_{23}\alpha_1+\underline{Y}_{23}\alpha_2,\
\Theta_{12}=\underline{T}_{22}+\underline{T}_{23}\beta_1+\underline{Y}_{23}\beta_2\nonumber\\
&&\Theta_{13}=\underline{Y}_{21}+\underline{T}_{23}
\gamma_1+\underline{Y}_{23} \gamma_2,\
\Theta_{14}=\underline{Y}_{22}+\underline{T}_{23}\delta_1+\underline{Y}_{23}\delta_2\nonumber\\
&&\Theta_{21}=-\underline{T}_{11}-\underline{T}_{13}\alpha_1-\underline{Y}_{13}\alpha_2.\
\Theta_{22}=-\mu_0 s\eta^{(3)}_{12}-\underline{T}_{13}\beta_1-\underline{Y}_{13}\beta_2,\nonumber\\
&&\Theta_{23}=-\underline{Y}_{11}-\underline{T}_{13}\gamma_1-\underline{Y}_{13}\gamma_2,\
\Theta_{24}= -\underline{Y}_{12}-\underline{T}_{13}\delta_1-\underline{Y}_{13}\delta_2 \nonumber\\
&&\Theta_{31}=\underline{Z}_{21}+\underline{Z}_{23}\alpha_1+\underline{W}_{23}\alpha_2.\
\Theta_{32}=\underline{Z}_{22}+\underline{Z}_{23}\beta_1+\underline{W}_{23}\beta_2\nonumber\\
&&\Theta_{33}=-
s\eta^{(2)}_{21}+\underline{Z}_{23}\gamma_1+\underline{W}_{23}\gamma_2,\
\Theta_{34}=
\underline{W}_{22}+\underline{Z}_{23}\delta_1+\underline{W}_{23}\delta_2\nonumber\\
&&\Theta_{41}=-\underline{Z}_{11}-\underline{Z}_{13}\alpha_1-\underline{W}_{13}\alpha_2,\
\Theta_{42}=-\underline{Z}_{12}-\underline{Z}_{13}\beta_1-\underline{W}_{13}\beta_2\nonumber\\
&&\Theta_{43}=-\underline{W}_{11}-\underline{Z}_{13}\gamma_1-\underline{W}_{13}\gamma_2,\
\Theta_{44}=-s\eta^{(2)}_{12}-\underline{Z}_{13}\delta_1-\underline{W}_{13}\delta_2
\end{eqnarray}
\begin{eqnarray}\label{c42.2}
G^{f,b}_1&=&\underline{J}^{f,b}_2+\frac{\underline{Z}_{33}\underline{Y}^{f,b}_{23}-\underline{T}_{23}\underline{W}_{33}}
{\underline{T}_{33}\underline{W}_{33}-\underline{Z}_{33}\underline{Y}_{33}}\
\underline{J}^{f,b}_3+\frac{\underline{T}_{23}\underline{Y}_{33}-\underline{Y}_{23}\underline{T}_{33}}{\underline{T}_{33}
\underline{W}_{33}-\underline{Z}_{33}\underline{Y}_{33}}\
\underline{J}^{f,b}_6\nonumber\\
\nonumber\\
G^{f,b}_2&=&-\underline{J}^{f,b}_1+\frac{\underline{T}_{13}\underline{W}_{33}-\underline{Y}_{13}\underline{Z}_{33}}
{\underline{T}_{33}\underline{W}_{33}-\underline{Z}_{33}\underline{Y}_{33}}\
\underline{J}^{f,b}_3+\frac{\underline{T}_{33}\underline{Y}_{13}-\underline{Y}_{33}\underline{T}_{13}}{\underline{T}_{33}
\underline{W}_{33}-\underline{Z}_{33}\underline{Y}_{33}}\
\underline{J}^{f,b}_6\nonumber\\
\nonumber\\
G^{f,b}_3&=&\underline{J}^{f,b}_5+\frac{\underline{W}_{23}\underline{Z}_{33}-\underline{Z}_{23}\underline{W}_{33}}
{\underline{T}_{33}\underline{W}_{33}-\underline{Z}_{33}\underline{Y}_{33}}\
\underline{J}^{f,b}_3+\frac{\underline{Z}_{23}\underline{Y}_{33}-\underline{W}_{23}\underline{T}_{33}}{\underline{T}_{33}
\underline{W}_{33}-\underline{Z}_{33}\underline{Y}_{33}}\
\underline{J}^{f,b}_6\nonumber\\
\nonumber\\
G^{f,b}_4&=&-\underline{J}^{f,b}_4+\frac{\underline{W}_{33}\underline{Z}_{13}-\underline{Z}_{33}\underline{W}_{13}}
{\underline{T}_{33}\underline{W}_{33}-\underline{Z}_{33}\underline{Y}_{33}}\
\underline{J}^{f,b}_3+\frac{\underline{W}_{13}\underline{T}_{33}-\underline{Z}_{13}\underline{Y}_{33}}{\underline{T}_{33}
\underline{W}_{33}-\underline{Z}_{33}\underline{Y}_{33}}\
\underline{J}^{f,b}_6\nonumber\\
&&
\end{eqnarray}
where $\underline{J}^{f,b}_i\ , i=1,2,3,4$ are the two dimensional
Fourier transforms of the operators $J^{f,b}_i\ , i=1,2,3,4$ and
$J^{f,b}_i\ , i=1,2,3,4$ are defined by (\ref{c32.1}).\\
 Since for a multilayer medium the susceptibility
 tensors are piecewisely constant, regarding to the definitions
 (\ref{c36}), (\ref{c37}),(\ref{c39}) and (\ref{c41.1}) the matrix $\Theta$ in
 (\ref{c40}) is independent of $z$ for each layer. Therefore Eq.(\ref{c40}) in a layer is a set of first
 order differential equations with constant coefficients. In this case the solution of Eq.(\ref{c40})
 in each layer is the sum of two parts
\begin{eqnarray}\label{c42}
\Lambda^{f,b}({\bf k}^\|,z,s)=\Lambda^{f,b}_g({\bf
k}^\|,z,s)+\Lambda^{f,b}_p({\bf k}^\|,z,s)
\end{eqnarray}
The part $\Lambda^{f,b}_p$ is a special solution for Eq.(\ref{c40})
that  can be written   as
\begin{eqnarray}\label{c43}
\Lambda^{f,b}_p({\bf k}^\|,s,z)=\frac{1}{2\pi}\int_D\ dz'
\int_{-\infty}^{+\infty} dq\ \frac{e^{iq(z-z')}}{iq I\pm\Theta({\bf
k}^\|,s)} G^{f,b}({\bf k}^\|,s,z')
\end{eqnarray}
where the integration domain $D$ is the width of the layer. The part
$\Lambda^{f,b}_g$ in (\ref{c42}) is the general solution of the
homogeneous equation corresponding to Eq.(\ref{c40}) and is as
\begin{eqnarray}\label{c44}
\Lambda^{f,b}_g({\bf k}^\|,z,s)=\sum_{j=1}^4\ C^{f,b}_j({\bf
k}^\|,s)\ R_j({\bf k}^\|,s)\exp[\mp\Omega_j({\bf k}^\|,s)z]
\end{eqnarray}
 where $R_j\ ,\ \Omega_j, j=1,2,3,4 $  are the eigenvectors  and the eigenvalues  of the matrix
 $\Theta$, respectively, that is
\begin{eqnarray}\label{c45}
\Theta({\bf k}^\|,s)R_j({\bf k}^\|,s)=\Omega_j({\bf k}^\|,s)
R_j({\bf k}^\|,s)\hspace{2.00 cm} j=1,2,3,4
\end{eqnarray}
 In (\ref{c44}) the operators $C^{f,b}_j,\ j=1,2,3,4$ should be obtained using the boundary conditions on the
 components of $\Lambda^{f,b}$. Regarding the definition of $\Lambda^{f,b}$ in (\ref{c41}) the components of
$\Lambda^{f,b}$ should be continuous on the boundaries between
layers. Also the components of $\Lambda^{f,b}$ should be
exponentially decay at $z\rightarrow\pm\infty$.\\
Consequently  using the obtained backward and forward Laplace
transformation of the four dimensional operator field $ \Lambda$, we
can express the time-space dependence  of $ \Lambda$ as
\begin{eqnarray}\label{c46}
&&\Lambda({\bf
r},t)=\frac{1}{(2\pi)^2}\int_{-\infty}^{+\infty}d\omega e^{-i\omega
t}\int_{-\infty}^{+\infty}d^2k[\Lambda^f({\bf k}^\|,
z,-i\omega+0^+)e^{i{\bf k}^\|\cdot{\bf
r}^\|}\nonumber\\
&&+\Lambda^b({\bf k}^\|, i\omega+0^+,z)e^{-i{\bf
k}^\|\cdot{\bf r}^\|}]\nonumber\\
&&
\end{eqnarray}
Finally substituting the components of the operator $\Lambda$ in
Eqs.(\ref{c38}) the z-components of the fields $E_z$ and $H_z$ can
be computed.
\section{A magnetodielectric slab}
Consider the region  $ 0<z<d$ to be filled by a homogeneous
magnetodielectric medium with susceptibilities tensors $
\chi^{(i)}(t)\   i=1,2,3,4$ and the regions $ z<0$ and $ z>d$ are
free space. That is for the regions $ z<0$ and $ z>d$ the
susceptibilities tensor $ \chi^{(i)}(t)\ i=1,2,3,4$ are identically
zero. Therefore according to the relations (\ref{c26})-(\ref{c29})
the tensors $ \eta^{(i)}\   i=1,2,3,4$ are zero for the regions $
z<0$ and $ z>d$. Using  the relations (\ref{c33}),
(\ref{c36}),(\ref{c37}),(\ref{c39}) and (\ref{c39.1}) for the free
spaces we deduce
\begin{eqnarray}\label{c47}
&&\alpha_1=\beta_1=\gamma_2=\delta_2=0\nonumber\\
&&\gamma_1=-\frac{ik_y}{s\varepsilon_0}\  , \
\delta_1=\frac{ik_x}{s\varepsilon_0}\ ,\
\alpha_2=\frac{ik_y}{s\mu_0}\ ,\ \beta_2=-\frac{ik_x}{s\mu_0}
\end{eqnarray}
Now from the definition of the matrix $\Theta$ in (\ref{c41.1}) and
using (\ref{c33}),(\ref{c36}), (\ref{c37}),(\ref{c39.1}),
(\ref{c41.1})and (\ref{c47}) this matrix for the regions $ z<0 $ and
$ z>d$ becomes as
\begin{eqnarray}\label{c48}
\Theta^{(0)}(\bf{k}^\| ,s)=\left[\begin{array}{cccc}
  0 & 0 & -\frac{k_x k_y}{s\varepsilon_0} & \mu_0s+\frac{k_x^2}{s\varepsilon_0} \\
  0 & 0 &  - \mu_0s-\frac{k_y^2}{s\varepsilon_0} & \frac{k_x k_y}{s\varepsilon_0} \\
   \frac{k_x k_y}{s\mu_0}& -\varepsilon_0s-\frac{k_x^2}{s\mu_0} & 0& 0 \\
   \varepsilon_0 s+\frac{k_y^2}{s \mu_0}& -\frac{k_xk_y}{s
        \mu_0} & 0 & 0 \\
\end{array}\right]
\end{eqnarray}
The eigenvalues of $\Theta^{(0)}(\bf{k}^\| ,s)$ are as
\begin{eqnarray}\label{c49}
&&\Omega^{(0)}_1(\bf{k}^\|,s)=\Omega^{(0)}_2(\bf{k}^\|,s)=-\sqrt{k_x^2+k_y^2+s^2\varepsilon_0
\mu_0}\nonumber\\
&&\Omega^{(0)}_3(\bf{k}^{\|},s)=\Omega^{(0)}_4(\bf{k}^\|,s)=\sqrt{k_x^2+k_y^2+s^2\varepsilon_0
\mu_0}
\end{eqnarray}
and corresponding eigenvectors are
\begin{eqnarray}\label{c50}
&&R^{(0)}_1=\left[\begin{array}{c}
                  -\frac{k_x^2+s^2\varepsilon_0\mu_0}{s\varepsilon_0\sqrt{k_x^2+k_y^2+s^2\varepsilon_0\mu_0}} \\
                  -\frac{k_x k_y}{s\varepsilon_0\sqrt{k_x^2+k_y^2+s^2\varepsilon_0\mu_0}} \\
                  0 \\
                  1 \\
                \end{array}\right]\hspace{1 cm}R^{(0)}_2=\left[\begin{array}{c}
                  \frac{k_x k_y}{s\varepsilon_0\sqrt{k_x^2+k_y^2+s^2\varepsilon_0\mu_0}} \\
                  \frac{k_y^2+s^2\varepsilon_0\mu_0}{s\varepsilon_0\sqrt{k_x^2+k_y^2+s^2\varepsilon_0\mu_0}} \\
                  1 \\
                  0 \\
                \end{array}\right]\nonumber\\
&&R^{(0)}_3=\left[\begin{array}{c}
                  \frac{k_x^2+s^2\varepsilon_0\mu_0}{s\varepsilon_0\sqrt{k_x^2+k_y^2+s^2\varepsilon_0\mu_0}} \\
                  \frac{k_x k_y}{s\varepsilon_0\sqrt{k_x^2+k_y^2+s^2\varepsilon_0\mu_0}} \\
                  0 \\
                  1 \\
                \end{array}\right]\hspace{1 cm}R^{(0)}_4=\left[\begin{array}{c}
                  -\frac{k_x k_y}{s\varepsilon_0\sqrt{k_x^2+k_y^2+s^2\varepsilon_0\mu_0}} \\
                  -\frac{k_y^2+s^2\varepsilon_0\mu_0}{s\varepsilon_0\sqrt{k_x^2+k_y^2+s^2\varepsilon_0\mu_0}} \\
                  1 \\
                  0 \\
                \end{array}\right]
\end{eqnarray}
For free spaces the coupling tensors $ \bf{f}_i , \bf{g}_i , i=1,2$
are zero and  from the definitions (\ref{c10}) , (\ref{c30}) and
(\ref{c31}) we deduce $ \bf{P}_N=\bf{P'}_N=\bf{M}_N=\bf{M'}_N=0$.
Therefore according to the relation (\ref{c32.1}) we have
\begin{equation}\label{c51}
\bf{J}^{f,b}=\left[\begin{array}{c}
                    \pm \bf{B}(\bf{r},0) \\
                    \mp \bf{D}(\bf{r},0)\\
                  \end{array}\right]
\end{equation}
 Using the relations (\ref{c33}), (\ref{c34}), (\ref{c36}), (\ref{c37}), (\ref{c39.1})and (\ref{c42.2})
 it is easy  to show that for free space the source terms $
 G^{f,b}_ i , i=1,2,3,4$ in (\ref{42.2})are reduced to
\begin{eqnarray}\label{c52}
&& G^{f,b}_1(\bf{k}^\|,s,z)=\pm\underline{
B}^{f,b}_y(\bf{k}^\|,z)\pm\frac{ik_x}{s\varepsilon_0}\underline{D}^{f,b}_z(\bf{k}^\|,z)\nonumber\\
&&G^{f,b}_2(\bf{k}^\|,s,z)=\mp\underline{
B}^{f,b}_x(\bf{k}^\|,z)\pm\frac{ik_y}{s\varepsilon_0}\underline{D}^{f,b}_z(\bf{k}^\|,z)\nonumber\\
&&G^{f,b}_3(\bf{k}^\|,s,z)=\mp\underline{
D}^{f,b}_y(\bf{k}^\|,z)\pm\frac{ik_x}{s\mu_0}\underline{B}^{f,b}_z(\bf{k}^\|,z)\nonumber\\
&&G^{f,b}_4(\bf{k}^\|,s,z)=\pm\underline{
D}^{f,b}_x(\bf{k}^\|,z)\pm\frac{ik_y}{s\mu_0}\underline{B}^{f,b}_z(\bf{k}^\|,z)
\end{eqnarray}
where
\begin{eqnarray}\label{c53}
&&\underline{\bf{D}}^{f,b}(\bf{k}^\|,z)=-i\sum_{\lambda=1}^2\int_{-\infty}^{+\infty}
dk_z \sqrt{\frac{\hbar\omega_{\bf{k}}} {4\pi}}\left(
a^\dag_{\mp\bf{k}\lambda}(0)\bf{e}_{\mp\bf{k}\lambda}-a^\dag_{\pm\bf{k}\lambda}(0)\bf{e}_{\pm\bf{k}\lambda}\right)e^{\pm
ik_z z }\nonumber\\
&&\underline{\bf{B}}^{f,b}(\bf{k}^\|,z)=
\sum_{\lambda=1}^2\int_{-\infty}^{+\infty}dk_z
\sqrt{\frac{\hbar}{4\pi\omega_{\bf{k}}}}(\pm i \bf{k}\times
\bf{e}_{\pm \bf{k}\lambda})a_{\pm\bf{k}\lambda}(0)\nonumber\\
&&+(\pm i \bf{k}\times
\bf{e}_{\mp\bf{k}\lambda})a^\dag_{\mp\bf{k}\lambda}(0)]e^{\pm ik_z
z}
\end{eqnarray}
Now using (\ref{c52}) and (\ref{c53}) it can be shown that the
special answer $ \Lambda^{f,b}_p$ in (\ref{c42})for free space is as
\begin{equation}\label{c54}
(\Lambda^{(0)}_p)^{f,b}(\bf{k}^\|,s,z)=\int dk_z \frac{e^{\pm i k_z
z}}{ \pm ik_z I\pm\Theta^{(0)}(\bf{k}^\|,s)} Q^{f,b}(\bf{k}^\|,k_z
,s)
\end{equation}
where $ Q^{f,b}$ is a matrix $4\times1$ with components
\begin{eqnarray}\label{c55}
&&Q^{f,b}_1(\bf{k}^\|,k_z ,s)=\nonumber\\
&&\pm\sum_{\lambda=1}^2 \sqrt{\frac{\hbar}{4\pi\omega_{\bf{k}}}}(\pm
i \bf{k}\times \bf{e}_{\pm
\bf{k}\lambda})_ya_{\pm\bf{k}\lambda}(0)+(\pm i \bf{k}\times
\bf{e}_{\mp\bf{k}\lambda})_ya^\dag_{\mp\bf{k}\lambda}(0)]\nonumber\\
&&\pm\frac{ik_x}{s\varepsilon_0}\left[-i\sum_{\lambda=1}^2
\sqrt{\frac{\hbar\omega_{\bf{k}}} {4\pi}}\left(
a^\dag_{\mp\bf{k}\lambda}(0)\bf{e}_{\mp\bf{k}\lambda z}-a^\dag_{\pm\bf{k}\lambda}(0)\bf{e}_{\pm\bf{k}\lambda z}\right)\right]\nonumber\\
&&Q^{f,b}_2(\bf{k}^\|,k_z ,s)=\nonumber\\
&&\mp\sum_{\lambda=1}^2 \sqrt{\frac{\hbar}{4\pi\omega_{\bf{k}}}}(\pm
i \bf{k}\times \bf{e}_{\pm
\bf{k}\lambda})_xa_{\pm\bf{k}\lambda}(0)+(\pm i \bf{k}\times
\bf{e}_{\mp\bf{k}\lambda})_xa^\dag_{\mp\bf{k}\lambda}(0)]\nonumber\\
&&\pm\frac{i k_y}{s\varepsilon_0}\left[-i\sum_{\lambda=1}^2
\sqrt{\frac{\hbar\omega_{\bf{k}}} {4\pi}}\left(
a^\dag_{\mp\bf{k}\lambda}(0)\bf{e}_{\mp\bf{k}\lambda z}-a^\dag_{\pm\bf{k}\lambda}(0)\bf{e}_{\pm\bf{k}\lambda z}\right)\right]\nonumber\\
&&Q^{f,b}_3(\bf{k}^\|,k_z ,s)=\mp\left[-i\sum_{\lambda=1}^2
\sqrt{\frac{\hbar\omega_{\bf{k}}} {4\pi}}\left(
a^\dag_{\mp\bf{k}\lambda}(0)\bf{e}_{\mp\bf{k}\lambda y}-a^\dag_{\pm\bf{k}\lambda}(0)\bf{e}_{\pm\bf{k}\lambda y}\right)\right]\nonumber\\
&&\pm\frac{i k_x}{\mu_0 s}[\sum_{\lambda=1}^2
\sqrt{\frac{\hbar}{4\pi\omega_{\bf{k}}}}(\pm i \bf{k}\times
\bf{e}_{\pm \bf{k}\lambda})_za_{\pm\bf{k}\lambda}(0) +(\pm i
\bf{k}\times
\bf{e}_{\mp\bf{k}\lambda})_za^\dag_{\mp\bf{k}\lambda}(0)]\nonumber\\
&&Q^{f,b}_4(\bf{k}^\|,k_z ,s)=\pm\left[-i\sum_{\lambda=1}^2
\sqrt{\frac{\hbar\omega_{\bf{k}}} {4\pi}}\left(
a^\dag_{\mp\bf{k}\lambda}(0)\bf{e}_{\mp\bf{k}\lambda x}-a^\dag_{\pm\bf{k}\lambda}(0)\bf{e}_{\pm\bf{k}\lambda x}\right)\right]\nonumber\\
&&\pm\frac{i k_y}{\mu_0 s}[\sum_{\lambda=1}^2
\sqrt{\frac{\hbar}{4\pi\omega_{\bf{k}}}}(\pm i \bf{k}\times
\bf{e}_{\pm \bf{k}\lambda})_za_{\pm\bf{k}\lambda}(0) +(\pm i
\bf{k}\times
\bf{e}_{\mp\bf{k}\lambda})_za^\dag_{\mp\bf{k}\lambda}(0)]\nonumber\\
\end{eqnarray}
According to (\ref{c42}) and (\ref{c44}) the solution of
differential equation (\ref{c40}) for the regions $z<0$ , $0<z<d$
and $z>d$ are respectively as
\begin{eqnarray}\label{c56}
&&(\Lambda^{(0)})^{f,b}(\bf{k}^\|,s,z)=\sum_{j=1}^4
C'^{f,b}_j(\bf{k}^\|,s) R^{(0)}_j(\bf{k}^\|,s)
e^{\mp\Omega^{(0)}_j(\bf{k}^\|,s)z}+(\Lambda^{(0)}_p)^{f,b}(\bf{k}^\|,s,z)\nonumber\\
&&\Lambda^{f,b}(\bf{k}^\|,s,z)=\sum_{j=1}^4 C^{f,b}_j(\bf{k}^\|,s)
R_j(\bf{k}^\|,s)
e^{\mp\Omega_j(\bf{k}^\|,s)z}+\Lambda_p^{f,b}(\bf{k}^\|,s,z)\nonumber\\
&&(\Lambda^{(0)})^{f,b}(\bf{k}^\|,s,z)=\sum_{j=1}^4
C''^{f,b}_j(\bf{k}^\|,s) R^{(0)}_j(\bf{k}^\|,s)
e^{\mp\Omega^{(0)}_j(\bf{k}^\|,s)z}+(\Lambda^{(0)}_p)^{f,b}(\bf{k}^\|,s,z)\nonumber\\
\end{eqnarray}
where $\Lambda_p^{f,b}$ is given by (\ref{c43}). Because
$\Lambda^{f,b}(\bf{k}^\|,s,z)$ should be tend to zero at $z=\pm
\infty$ from the eigenvalues (\ref{c49})we deduce
\begin{eqnarray}\label{c57}
&& C'^f_3=C'^f_4=C'^b_1=C'^b_2=0\nonumber\\
&&C''^b_3=C''^b_4=C''^f_1=C''^f_2=0
\end{eqnarray}
The boundary condition at $z=0$ give us
\begin{eqnarray}\label{c58}
&&C'^f_1(\bf{k}^\|,s)R^{(0)}_1(\bf{k}^\|,s)+C'^f_2(\bf{k}^\|,s)R^{(0)}_2(\bf{k}^\|,s)+(\Lambda^{(0)}_p)^{f}(\bf{k}^\|,s,0)\nonumber\\
&&=\sum_{j=1}^4 C^{f}_j(\bf{k}^\|,s) R_j(\bf{k}^\|,s)
+\Lambda_p^{f}(\bf{k}^\|,s,0)\nonumber\\
&&C'^b_3(\bf{k}^\|,s)R^{(0)}_3(\bf{k}^\|,s)+C'^b_4(\bf{k}^\|,s)R^{(0)}_4(\bf{k}^\|,s)+(\Lambda^{(0)}_p)^{b}(\bf{k}^\|,s,0)\nonumber\\
&&=\sum_{j=1}^4 C^{b}_j(\bf{k}^\|,s) R_j(\bf{k}^\|,s)
+\Lambda_p^{b}(\bf{k}^\|,s,0)
\end{eqnarray}
Also the boundary condition at $z=d$ implies
\begin{eqnarray}\label{c59}
&&C''^f_3(\bf{k}^\|,s)R^{(0)}_3(\bf{k}^\|,s)e^{-\Omega^{(0)}_3(\bf{k}^\|,s)d}+C''^f_4(\bf{k}^\|,s)R^{(0)}_4(\bf{k}^\|,s)e^{-\Omega^{(0)}_4(\bf{k}^\|,s)d}\nonumber\\
&&+(\Lambda^{(0)}_p)^{f}(\bf{k}^\|,s,d)=\sum_{j=1}^4
C^{f}_j(\bf{k}^\|,s) R_j(\bf{k}^\|,s)e^{-\Omega_j(\bf{k}^\|,s)d}
+\Lambda_p^{f}(\bf{k}^\|,s,d)\nonumber\\
&&C''^b_1(\bf{k}^\|,s)R^{(0)}_1(\bf{k}^\|,s)e^{\Omega^{(0)}_1(\bf{k}^\|,s)d}+C''^b_2(\bf{k}^\|,s)R^{(0)}_2(\bf{k}^\|,s)e^{\Omega^{(0)}_2(\bf{k}^\|,s)d}\nonumber\\
&&+(\Lambda^{(0)}_p)^{b}(\bf{k}^\|,s,d)=\sum_{j=1}^4
C^{b}_j(\bf{k}^\|,s) R_j(\bf{k}^\|,s)e^{\Omega_j(\bf{k}^\|,s)d}
+\Lambda_p^{b}(\bf{k}^\|,s,d)\nonumber\\
\end{eqnarray}
Using the relations (\ref{c58}) we can obtain the coeffi-\\
cients $C^{f,b}(\bf{k}^\|,s)\ i=1,2,3,4$ in terms of $ C'^f_1,
C'^f_2, C'^b_3$ and $ C'^b_4$ as
\begin{eqnarray}\label{c60}
&&\left[\begin{array}{c}
          C^f_1 \\
          C^f_2 \\
          C^f_3 \\
          C^f_4 \\
        \end{array}\right]=C'^f_1
        \Gamma^{-1}(\bf{k}^\|,s)R^{(0)}_1(\bf{k}^\|,s)+C'^f_2
        \Gamma^{-1}(\bf{k}^\|,s)R^{(0)}_2(\bf{k}^\|,s)\nonumber\\
&&+\Gamma^{-1}(\bf{k}^\|,s)\left((\Lambda^{(0)}_p)^f(\bf{k}^\|,s,0)-\Lambda^f_p(\bf{k}^\|,s,0)\right)
\end{eqnarray}
\begin{eqnarray}\label{c61}
&&\left[\begin{array}{c}
          C^b_1 \\
          C^b_2 \\
          C^b_3 \\
          C^b_4 \\
        \end{array}\right]=C'^b_3
        \Gamma^{-1}(\bf{k}^\|,s)R^{(0)}_3(\bf{k}^\|,s)+C'^b_4
        \Gamma^{-1}(\bf{k}^\|,s)R^{(0)}_4(\bf{k}^\|,s)\nonumber\\
&&+\Gamma^{-1}(\bf{k}^\|,s)\left((\Lambda^{(0)}_p)^b(\bf{k}^\|,s,0)-\Lambda^b_p(\bf{k}^\|,s,0)\right)
\end{eqnarray}
where $\Gamma(\bf{k}^\|,s)$ is a $4\times4$ matrix are given by
eigenvectors $ R_i(\bf{k}^\|,s) , i=1,2,3,4$ as the following
\begin{equation}\label{c62}
\Gamma(\bf{k}^\|,s)=\left[\begin{array}{cccc}
                            R_1(\bf{k}^\|,s) & R_2(\bf{k}^\|,s) & R_3(\bf{k}^\|,s) & R_4(\bf{k}^\|,s) \\
                          \end{array}\right]
\end{equation}
Also using Eq. (\ref{c59}) it is clear that one can write the
coefficients $ C^{f,b}_i\    i=1,2,3,4$ in terms of $ C''^f_3,
C''^f_4, C''^b_1$ and $C''^b_2$ as
\begin{eqnarray}\label{c63}
&&\left[\begin{array}{c}
          C^f_1 \\
          C^f_2 \\
          C^f_3 \\
          C^f_4 \\
        \end{array}\right]=C''^f_3
        \triangle^{-1}(\bf{k}^\|,s)R^{(0)}_3(\bf{k}^\|,s)e^{-\Omega^{(0)}_3(\bf{k}^\|,s)d}\nonumber\\
&&+C''^f_4\triangle^{-1}(\bf{k}^\|,s)R^{(0)}_4(\bf{k}^\|,s)e^{-\Omega^{(0)}_4(\bf{k}^\|,s)d}\nonumber\\
&&+\triangle^{-1}(\bf{k}^\|,s)\left((\Lambda^{(0)}_p)^f(\bf{k}^\|,s,d)-\Lambda^f_p(\bf{k}^\|,s,d)\right)
\end{eqnarray}
\begin{eqnarray}\label{c64}
&&\left[\begin{array}{c}
          C^b_1 \\
          C^b_2 \\
          C^b_3 \\
          C^b_4 \\
        \end{array}\right]=C''^b_1
        \Pi^{-1}(\bf{k}^\|,s)R^{(0)}_1(\bf{k}^\|,s)e^{\Omega^{(0)}_1(\bf{k}^\|,s)d}\nonumber\\
&&+C''^b_2 \Pi^{-1}(\bf{k}^\|,s)R^{(0)}_2(\bf{k}^\|,s)e^{\Omega^{(0)}_2(\bf{k}^\|,s)d}\nonumber\\
&&+\Pi^{-1}(\bf{k}^\|,s)\left((\Lambda^{(0)}_p)^b(\bf{k}^\|,s,d)-\Lambda^b_p(\bf{k}^\|,s,d)\right)
\end{eqnarray}
where $\triangle^{-1}(\bf{k}^\|,s)$  and $\Pi^{-1}(\bf{k}^\|,s)$ are
$4\times4$ matrices  defined by
\begin{eqnarray}\label{c65}
&&\triangle^{-1}=\left[\begin{array}{cccc}
                                      R_1e^{-\Omega_1d} & R_2e^{-\Omega_2d}
                                      & R_3e^{-\Omega_3d} & R_4e^{-\Omega_4d} \\
                                    \end{array}\right]\nonumber\\
&&\Pi^{-1}=\left[\begin{array}{cccc}
                                      R_1e^{\Omega_1d} & R_2e^{\Omega_2d}
                                      & R_3e^{\Omega_3d} & R_4e^{\Omega_4d} \\
                                    \end{array}\right]
\end{eqnarray}
 from Eqs.(\ref{c60}) and (\ref{c63}) we deduce
\begin{eqnarray}\label{c66}
&&C'^f_1
        \Gamma^{-1}(\bf{k}^\|,s)R^{(0)}_1(\bf{k}^\|,s)+C'^f_2
        \Gamma^{-1}(\bf{k}^\|,s)R^{(0)}_2(\bf{k}^\|,s)\nonumber\\
&&+\Gamma^{-1}(\bf{k}^\|,s)\left((\Lambda^{(0)}_p)^f(\bf{k}^\|,s,0)-\Lambda^f_p(\bf{k}^\|,s,0)\right)\nonumber\\
&&=C''^f_3\triangle^{-1}(\bf{k}^\|,s)R^{(0)}_3(\bf{k}^\|,s)e^{-\Omega^{(0)}_3(\bf{k}^\|,s)d}\nonumber\\
&&+C''^f_4\triangle^{-1}(\bf{k}^\|,s)R^{(0)}_4(\bf{k}^\|,s)e^{-\Omega^{(0)}_4(\bf{k}^\|,s)d}\nonumber\\
&&+\triangle^{-1}(\bf{k}^\|,s)\left((\Lambda^{(0)}_p)^f(\bf{k}^\|,s,d)-\Lambda^f_p(\bf{k}^\|,s,d)\right)
\end{eqnarray}
where relate the coefficients $C'^f_1, C'^f_2$ in region $ z<0 $ to
the coefficients $ C''^f_3 , C''^f_4$ in region $ z>d $.  Using
Eqs.(\ref{c61}) and (\ref{c64}) we can write
\begin{eqnarray}\label{c67}
&&C'^b_3 \Gamma^{-1}(\bf{k}^\|,s)R^{(0)}_3(\bf{k}^\|,s)+C'^b_4
        \Gamma^{-1}(\bf{k}^\|,s)R^{(0)}_4(\bf{k}^\|,s)\nonumber\\
&&+\Gamma^{-1}(\bf{k}^\|,s)\left((\Lambda^{(0)}_p)^b(\bf{k}^\|,s,0)-\Lambda^b_p(\bf{k}^\|,s,0)\right)\nonumber\\
&&=C''^b_1\Pi^{-1}(\bf{k}^\|,s)R^{(0)}_1(\bf{k}^\|,s)e^{\Omega^{(0)}_1(\bf{k}^\|,s)d}\nonumber\\
&&+C''^b_2 \Pi^{-1}(\bf{k}^\|,s)R^{(0)}_2(\bf{k}^\|,s)e^{\Omega^{(0)}_2(\bf{k}^\|,s)d}\nonumber\\
&&+\Pi^{-1}(\bf{k}^\|,s)\left((\Lambda^{(0)}_p)^b(\bf{k}^\|,s,d)-\Lambda^b_p(\bf{k}^\|,s,d)\right)
\end{eqnarray}
where relate the coefficients $C'^b_3, C'^b_4$ in region $ z<0 $ to
the coefficients $ C''^b_1 , C''^b_2$ in region $ z>d $.Equation
(\ref{c66}) is a frame of algebraic equations which one can find the
coefficients  $ C'^f_1, C'^f_2 , C''^f_3  , C''^f_4$ using this
frame. Then, by insertion these coefficients in Eqs.(\ref{c60}) or
(\ref{c63}) one can calculate the coefficients $C^f_i\ i=1,2,3,4$.
Also by solving  the algebraic equations (\ref{c67}) the
coefficients $C'^b_3 , C'^b_4 , C''^b_1 , C''^b_2$ can be fined and
by instituting them in (\ref{c61}) or ((\ref{c64})the coefficients $
C^b_i\ i=1,2,3,4 $ can be computed.
\section{Summary and conclusion}
A bi-anisotropic magnetodielectric medium was modeled by two
independent reservoirs. Each reservoir contains a continuous set of
three dimensional harmonic oscillators. In contrast of the damped
polarization model, it is not needed the electric and magnetic
polarization fields of the medium to appear in the Lagrangian of the
total system and the reservoirs solely constitute the degrees of
freedom of the medium. The electric and magnetic polarization fields
of the medium were obtained in terms of the dynamical variable
modeling the medium and the coupling tensors of the medium and
electromagnetic field. The constitutive relation of the
bi-anisotropic magnetodielectric medium was obtained as a
consequence of the Euler-Lagrange equations of the total system. For
a multilayer medium, combination of the Maxwell equations and the
constitutive relation of the medium was leaded to a frame of first
order differential equations in terms of the electric and magnetic
fields. By solving the differential equations in a standard way the
electric and magnetic fields were computed for a bi-anisotropic
magnetodielectric  multilayer medium.

\end{document}